\documentclass[aps,pra,preprintnumbers,showpacs,tightenlines]{revtex4}

\usepackage{amssymb}
\usepackage{amsmath}
\usepackage{graphicx}
\usepackage{epsfig}
\usepackage{subfigure}
\usepackage{amsfonts}
\usepackage{CJK}

\begin{document}

\title{Fast quantum information transfer with superconducting flux qubits
coupled to a cavity}

\author{Chui-Ping Yang}
\email{yangcp@hznu.edu.cn}
\address{Department of Physics, Hangzhou Normal
University, Hangzhou, Zhejiang 310036, China}

\address{State Key Laboratory of Precision Spectroscopy, Department of Physics,
East China Normal University, Shanghai 200062, China}

\date{\today}

\begin{abstract}
We present a way to realize quantum information transfer with
superconducting flux qubits coupled to a cavity. Because only
resonant qubit-cavity interaction and resonant
qubit-pulse interaction are applied, the information transfer can be
performed much faster, when compared with the previous proposals. This
proposal does not require adjustment of the qubit level spacings during the
operation. Moreover, neither uniformity in the device parameters nor exact
placement of qubits in the cavity is needed by this proposal.

\end{abstract}

\maketitle
\date{\today}

\begin{center}
\textbf{I. INTRODUCTION}
\end{center}

The physical system, composed of circuit cavities and superconducting qubits
(such as charge, phase and flux qubits), has appeared to be one of the most
promising candidates for realizing scalable quantum information processing.
Superconducting qubits and microwave cavities can be fabricated with modern
integrated circuit technology, a superconducting qubit has relatively long
decoherence time [1,2], and a superconducting microwave cavity or resonator
acts as a ``quantum bus'', which can mediate long-range and fast interaction
between distant superconducting qubits [3-5]. Moreover, the strong coupling
between the cavity field and superconducting qubits, which is difficult to
achieve with atoms in a microwave cavity, was earlier predicted by theory
[6,7] and has been experimentally demonstrated [8,9].

On the other hand, much attention has been paid to quantum information
transfer (QIT). One example to illustrate the importance of QIT is as
follows. When performing quantum information processing in a practical
system, one needs to transfer the state of the operation qubit to the memory
qubit for storage after a step of processing is completed; and one needs to
transfer the state from the memory qubit back to the operation qubit when a
further step of processing is needed. Within cavity QED technique, QIT has
been experimentally demonstrated with superconducting phase qubits and
transmon qubits in cavity QED [4,10]. However, to the best of our knowledge,
no experimental demonstration of QIT with superconducting flux qubits in
cavity QED has been reported.

Theoretical methods for implementing QIT [3,6,11-14] have been presented
with flux qubits (e.g., SQUID qubits) or charge-flux qubits based on cavity
QED technique. However, these methods have some drawbacks. For instances:
(i) the method presented in [3] requires adjustment of the level spacings of
the devices during the operation; (ii) the methods proposed in [6,11-13]
require slowly changing the Rabi frequencies to satisfy the adiabatic
passage; and (iii) the approach introduced in [14] requires a second-order
detuning to achieve an off-resonant Raman coupling between two relevant
levels. Note that the adjustment of the level spacings during the operation
is undesirable and also may cause extra decoherence. In addition, when the
adiabatic passage or a second-order detuning is applied, the operation
becomes slow (the operation time required for the information transfer is on
the order of one microsecond to a few microseconds [6,14]).

In this paper, we propose an alternative method for realizing QIT with
four-level superconducting flux qubits coupled to a cavity or resonator.
This proposal has the advantages: (i) because only resonant interactions are
applied, the speed of the operation is increased by two orders of magnitude
(as shown below, the operation time is on the order of $\sim $1 ns), when
compared with the previous proposals [6,11-14] employing a second-order
large detuning or adiabatic passage; (ii) the method does not need
adjustment of the qubit level spacings during the operation, and thus
decoherence caused due to the adjustment of the qubit level spacings is
avoided; and (iii) the qubit-cavity coupling constants are not required to
be identical for each qubit, therefore superconducting devices, which often
have considerable parameter nonuniformity, can be used and exact placement
of qubits in the cavity is not required.

This paper is organized as follows. In Sec.~II, we briefly review the basic
theory of resonant qubit-cavity and qubit-pulse interactions. In Sec.~III,
we show how to realize QIT with superconducting flux qubits coupled to a
cavity or resonator. In Sec.~IV, we briefly discuss possible experimental
implementation with superconducting flux qubits coupled to a one-dimensional
transmission line resonator. A concluding summary is presented in Sec. V.

\begin{figure}[tbp]
\includegraphics[bb=82 362 265 539, width=4.6 cm, clip]{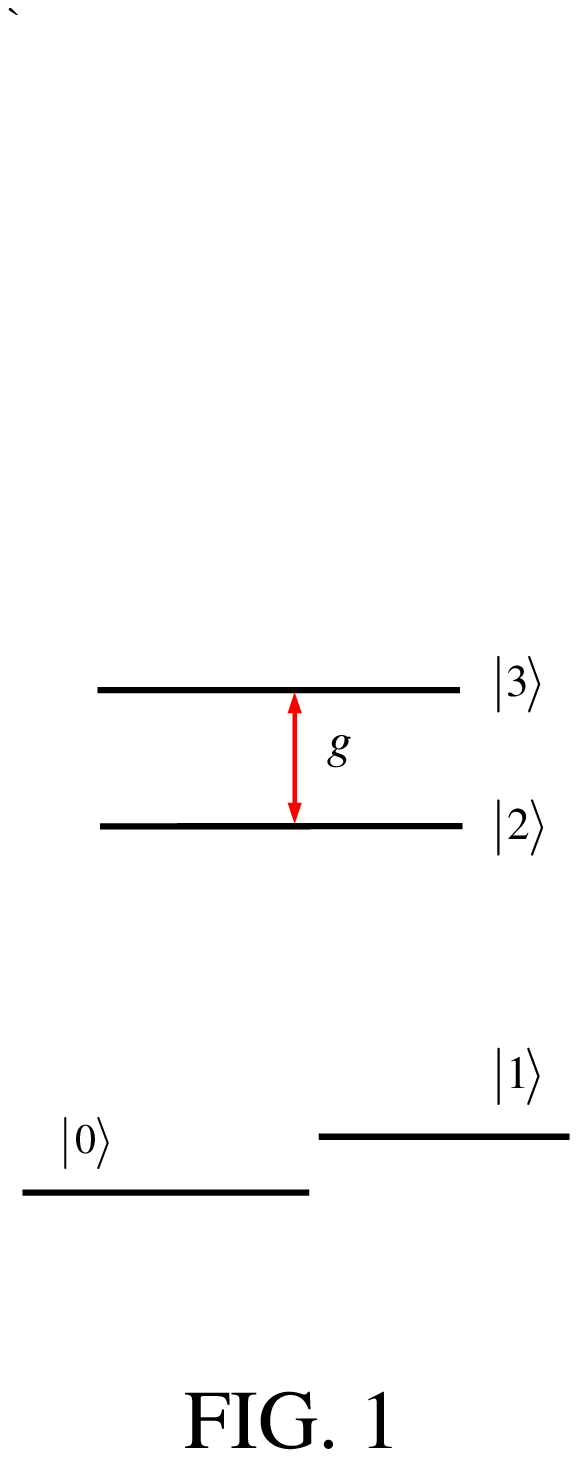} %
\vspace*{-0.08in}
\caption{(Color online) Level diagram of a four-level flux qubit, with
forbidden or weak transition between the two lowest levels. The cavity mode
is resonant with the transition between the top two levels. $g$ is the
coupling constant between the cavity mode and the $\left| 2\right\rangle
\leftrightarrow \left| 3\right\rangle $ transition.}
\label{fig:1}
\end{figure}

\begin{center}
\textbf{II. BASIC THEORY}
\end{center}

The flux qubits throughout this paper have four levels $\left|
0\right\rangle ,$ $\left| 1\right\rangle ,$ $\left| 2\right\rangle ,$ and $%
\left| 3\right\rangle $ as depicted in Fig. 1. In general, there exists the
transition between the two lowest levels $\left| 0\right\rangle $ and $%
\left| 1\right\rangle $, which however can be made to be weak via increasing
the potential barrier between the two levels $\left| 0\right\rangle $ and $%
\left| 1\right\rangle $ [1,15,16]. The qubits with this four-level structure
could be a radio-frequency superconducting quantum interference device (rf
SQUID) consisting of one Josephson junction enclosed by a superconducting
loop, or a superconducting device with three Josephson junctions enclosed by
a superconducting loop. For flux qubits, the two logic states of a qubit are
represented by the two lowest levels $\left| 0\right\rangle $ and $\left|
1\right\rangle .$

\begin{center}
\textbf{A. Qubit-cavity resonant interaction}
\end{center}

Consider a flux qubit with four levels as shown in Fig.~1. Suppose that the
transition between the two levels $\left| 2\right\rangle $ and $\left|
3\right\rangle $ is resonant with the cavity mode. In the interaction
picture and under the rotating-wave approximation, the interaction
Hamiltonian of the qubit and the cavity mode is given by
\begin{equation}
H=\hbar g(a^{+}\sigma _{23}^{-}+\text{H.c.}),
\end{equation}
where $a^{+}$ and $a$ are the photon creation and annihilation operators of
the cavity mode, $g$ is the coupling constant between the cavity mode and
the $\left| 2\right\rangle \leftrightarrow \left| 3\right\rangle $
transition of the qubit, and $\sigma _{23}^{-}=\left| 2\right\rangle
\left\langle 3\right| $.

Based on the Hamiltonian (1), it can be easily found that the initial states
$\left| 3\right\rangle \left| 0\right\rangle _c$ and $\left| 2\right\rangle
\left| 1\right\rangle _c$ of the qubit and the cavity mode evolve as follows
\begin{eqnarray}
\left| 3\right\rangle \left| 0\right\rangle _c &\rightarrow &-i\sin \left(
gt\right) \left| 2\right\rangle \left| 1\right\rangle _c+\cos (gt)\left|
3\right\rangle \left| 0\right\rangle _c,  \nonumber \\
\left| 2\right\rangle \left| 1\right\rangle _c &\rightarrow &\cos \left(
gt\right) \left| 2\right\rangle \left| 1\right\rangle _c-i\sin \left(
gt\right) \left| 3\right\rangle \left| 0\right\rangle _c.
\end{eqnarray}
However, the state $\left| 0\right\rangle \left| 0\right\rangle _c$ remains
unchanged under the Hamiltonian (1).

The coupling strength $g$ may vary with different qubits due to non-uniform
device parameters and/or non-exact placement of qubits in the cavity.
Therefore, in the operation below, $g$ will be replaced by $g_1$ and $g_2$
for qubits $1$ and $2$, respectively.

\begin{center}
\textbf{B. Qubit-pulse resonant interaction}
\end{center}

Consider a flux qubit with four levels as depicted in Fig.~1, driven by a
classical pulse. Suppose that the cavity mode is resonant with the
transition between the two levels $\left| i\right\rangle $ and $\left|
j\right\rangle $ of the qubit. Here, the level $\left| i\right\rangle $ is
the lower energy level. In the interaction picture and under the
rotating-wave approximation, the interaction Hamiltonian is given by
\begin{equation}
H_I=\hbar \left( \Omega _{ij}e^{i\phi }\left| i\right\rangle \left\langle
j\right| +\text{H.c.}\right) ,
\end{equation}
where $\Omega _{ij}$ is the Rabi frequency of the pulse and $\phi $ are the
initial phase of the pulse. Based on the Hamiltonian (3), it is
straightforward to show that a pulse of duration $t$ results in the
following state transformation
\begin{eqnarray}
\left| i\right\rangle &\rightarrow &\cos \Omega _{ij}t\left| i\right\rangle
-ie^{-i\phi }\sin \Omega _{ij}t\left| j\right\rangle ,  \nonumber \\
\left| j\right\rangle &\rightarrow &\cos \Omega _{ij}t\left| j\right\rangle
-ie^{i\phi }\sin \Omega _{ij}t\left| i\right\rangle ,
\end{eqnarray}
which can be completed within a very short time, by increasing the pulse
Rabi frequency $\Omega _{ij}$ (i.e., by increasing the intensity of the
pulse).

In above we have given a discussion on the qubit-cavity resonant interaction
and the qubit-pulse resonant interaction. The results (2) and (4) presented
above will be used for the QIT implementation below.

\begin{center}
\textbf{III. REALIZING QIT WITH FLUX QUBITS IN CAVITY QED}
\end{center}

Consider two flux qubits 1 and 2. Each qubit has a four-level configuration
as depicted in Fig.~1. To begin with, it should be mentioned that during the
operations below, the following conditions are required, which are: (i) the
cavity mode is resonant with the $\left| 2\right\rangle \leftrightarrow
\left| 3\right\rangle $ transition of each qubit, (ii) the cavity mode is
highly detuned (decoupled) from the transition between any other two levels,
and (iii) the pulse is resonant with the transition between two relevant
levels of each qubit but highly detuned (decoupled) from the transition
between any two irrelevant levels of each qubit. The first condition can be
achieved by setting the level spacing between the two levels $\left|
2\right\rangle $ and $\left| 3\right\rangle $ to be the same for each qubit.
Note that for superconducting qubits, by designing the qubits appropriately,
one can easily make the level spacing between certain two levels (the two
levels $\left| 2\right\rangle $ and $\left| 3\right\rangle $ here) to be
identical [17], though it is hard to have the level spacing between any two
levels to be identical for each qubit due to nonuniformity of the device
parameters. In addition, the second and third conditions can be achieved via
prior adjustment of the qubit level spacings before the operation. For
superconducting flux qubits, the level spacings can be readily adjusted by
changing the external flux applied to the superconducting loop
[1,15,16,18,19]). With these in mind, we now give a detailed discussion on
how to realize the QIT.

\begin{figure}[tbp]
\includegraphics[bb=53 223 593 655, width=9.6 cm, clip]{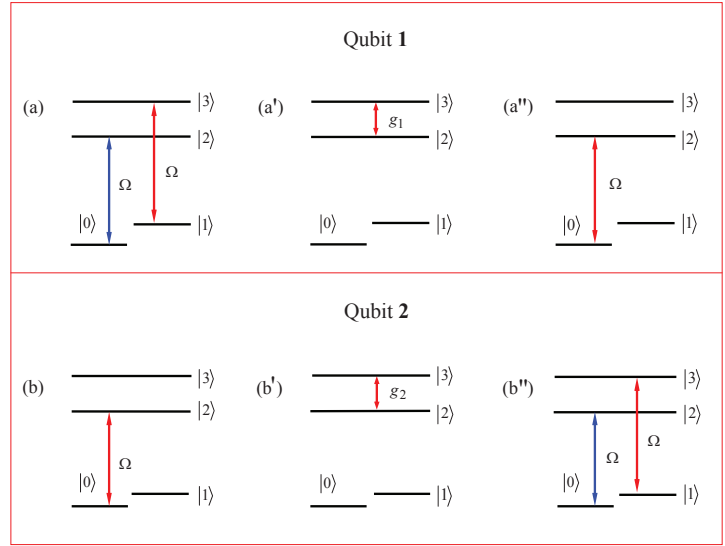} %
\vspace*{-0.08in}
\caption{(Color online) Illustration of qubit 1 or qubit 2 interacting with
the cavity mode or the pulses during the QIT.}
\label{fig:2}
\end{figure}

The cavity mode is initially in the vacuum state $\left| 0\right\rangle _c.$
The procedure for realizing QIT is listed below:

Step (i): (a) Apply a pulse (with a frequency $\omega =\omega _{31},$ a
phase $\phi =\pi $) and a pulse (with a frequency $\omega =\omega _{20},$ a
phase $\phi =-\frac \pi 2$) to qubit 1 [Fig.~2(a)]; the duration of each
pulse is $t_{1,a}=\frac \pi {2\Omega };$ according to Eq.~(4), the first
pulse leads to $\left| 1\right\rangle _1\rightarrow $ $i\left|
3\right\rangle _1$ while the second pulse results in $\left| 0\right\rangle
_1\rightarrow $ $\left| 2\right\rangle _1;$ (b) Wait for a time $%
t_{1,b}=\frac \pi {2g_1}$ to have the cavity mode resonantly interacting
with the $\left| 2\right\rangle \leftrightarrow \left| 3\right\rangle $
transition of qubit $1$ [Fig.~2(a$^{\prime }$)], resulting in $\left|
3\right\rangle _1\left| 0\right\rangle _c\rightarrow $ $-i\left|
2\right\rangle _1\left| 1\right\rangle _c$ as described by Eq. (2) while
nothing to the state $\left| 2\right\rangle _1\left| 0\right\rangle _c$;
then (c) Apply a pulse (with a frequency $\omega =\omega _{20},$ a phase $%
\phi =\frac \pi 2,$ and a duration $t_{1,c}=\frac \pi {2\Omega }$) to qubit $%
1$ [Fig.~2(a$^{\prime \prime }$)], resulting in $\left| 2\right\rangle
_1\rightarrow \left| 0\right\rangle _1$. It can be seen that after the
operation of this step, the following transformation is obtained:

\begin{equation}
\begin{array}{c}
\left| 0\right\rangle _1\left| 0\right\rangle _c\otimes \left|
0\right\rangle _2 \\
\left| 1\right\rangle _1\left| 0\right\rangle _c\otimes \left|
0\right\rangle _2
\end{array}
\stackrel{\left( a\right) }{\rightarrow }
\begin{array}{c}
\left| 2\right\rangle _1\left| 0\right\rangle _c\otimes \left|
0\right\rangle _2 \\
i\left| 3\right\rangle _1\left| 0\right\rangle _c\otimes \left|
0\right\rangle _2
\end{array}
\stackrel{\left( b\right) }{\rightarrow }
\begin{array}{c}
\left| 2\right\rangle _1\left| 0\right\rangle _c\otimes \left|
0\right\rangle _2\stackrel{\left( c\right) }{\rightarrow } \\
\left| 2\right\rangle _1\left| 1\right\rangle _c\otimes \left|
0\right\rangle _2
\end{array}
\begin{array}{c}
\left| 0\right\rangle _1\left| 0\right\rangle _c\otimes \left|
0\right\rangle _2 \\
\left| 0\right\rangle _1\left| 1\right\rangle _c\otimes \left|
0\right\rangle _2
\end{array}
.
\end{equation}

Step (ii): (a) Apply a pulse (with a frequency $\omega =\omega _{20},$ a
phase $\phi =-\frac \pi 2,$ and a duration $t_{2,a}=\frac \pi {2\Omega }$)
to qubit $2$ [Fig.~2(b)], to transform $\left| 0\right\rangle _2$ to $\left|
2\right\rangle _2$; (b) Wait for a time $t_{2,b}=\frac \pi {2g_2}$ to have
the cavity mode resonantly interacting with the $\left| 2\right\rangle
\leftrightarrow \left| 3\right\rangle $ transition of qubit $2$ [Fig.~2(b$%
^{\prime }$)], such that the state $\left| 2\right\rangle _2\left|
1\right\rangle _c$ is transformed to $-i\left| 3\right\rangle _2\left|
0\right\rangle _c$; then (c) Apply a pulse (with a frequency $\omega =\omega
_{31},$ a phase $\phi =\pi $) and a pulse (with a frequency $\omega =\omega
_{20},$ a phase $\phi =\frac \pi 2$) qubit $1$ [Fig.~2(b$^{\prime \prime }$%
)]; the two pulses have the same duration $t_{2,c}=\frac \pi {2\Omega };$
the first pulse transforms the state $\left| 3\right\rangle _2$ to $i\left|
1\right\rangle _2$ while the second pulse transforms the state $\left|
2\right\rangle _2$ to $\left| 0\right\rangle _2.$

It can be seen that the operation of this step results in the following
transformation:

\begin{equation}
\begin{array}{c}
\left| 0\right\rangle _1\otimes \left| 0\right\rangle _2\left|
0\right\rangle _c \\
\left| 0\right\rangle _1\otimes \left| 0\right\rangle _2\left|
1\right\rangle _c
\end{array}
\stackrel{\left( a\right) }{\rightarrow }
\begin{array}{c}
\left| 0\right\rangle _1\otimes \left| 2\right\rangle _2\left|
0\right\rangle _c \\
\left| 0\right\rangle _1\otimes \left| 2\right\rangle _2\left|
1\right\rangle _c
\end{array}
\stackrel{\left( b\right) }{\rightarrow }
\begin{array}{c}
\left| 0\right\rangle _1\otimes \left| 2\right\rangle _2\left|
0\right\rangle _c \\
\left| 0\right\rangle _1\otimes \left( -i\right) \left| 3\right\rangle
_2\left| 0\right\rangle _c
\end{array}
\stackrel{\left( c\right) }{\rightarrow }
\begin{array}{c}
\left| 0\right\rangle _1\otimes \left| 0\right\rangle _2\left|
0\right\rangle _c \\
\left| 0\right\rangle _1\otimes \left| 1\right\rangle _2\left|
0\right\rangle _c
\end{array}
\end{equation}

Eq.~(5) shows that during the operations of step (i) on qubit 1 and the
cavity, the states $\left| 0\right\rangle _2$ and $\left| 1\right\rangle _2$
of qubit 2 do not change. In addition, Eq. (6) shows that during the
operation of step (ii) on qubit 2 and the cavity, the states $\left|
0\right\rangle _1$ and $\left| 1\right\rangle _1$ of qubit 1 remain
unchanged. This is because the cavity mode was initially assumed to be
resonant with the $\left| 2\right\rangle \leftrightarrow \left|
3\right\rangle $ transition but highly detuned (decoupled ) from the
transition between any other two levels of each qubit.

Based on the results (5) and (6), we obtain the transformation below:

\begin{equation}
\begin{array}{c}
\left| 0\right\rangle _1\left| 0\right\rangle _2\left| 0\right\rangle _c \\
\left| 1\right\rangle _1\left| 0\right\rangle _2\left| 0\right\rangle _c
\end{array}
\stackrel{\text{Step(i)}}{\longrightarrow }
\begin{array}{c}
\left| 0\right\rangle _1\left| 0\right\rangle _2\left| 0\right\rangle _c \\
\left| 0\right\rangle _1\left| 0\right\rangle _2\left| 1\right\rangle _c
\end{array}
\stackrel{\text{Step(ii)}}{\longrightarrow }
\begin{array}{c}
\left| 0\right\rangle _1\left| 0\right\rangle _2\left| 0\right\rangle _c \\
\left| 0\right\rangle _1\left| 1\right\rangle _2\left| 0\right\rangle _c
\end{array}
.
\end{equation}
From Eq.~(7), it is easy to see that when qubit 1 is initially in the state $%
\alpha \left| 0\right\rangle _1+\beta \left| 1\right\rangle _1$ (with $%
\left| \alpha \right| ^2+\left| \beta \right| ^2=1$) and qubit 2 initially
in the state $\left| 0\right\rangle _2$ before the operation of step (i),
the initial state $(\alpha \left| 0\right\rangle _1+\beta \left|
1\right\rangle _1)\left| 0\right\rangle _2\left| 0\right\rangle _c$ of the
whole system is transformed to the state $\left| 0\right\rangle _1(\alpha
\left| 0\right\rangle _2+\beta \left| 1\right\rangle _2)\left|
0\right\rangle _c$ after the above two-step operation. This result implies
that after the above operations, the cavity mode returns to its original
vacuum state; while the following transformation
\begin{equation}
(\alpha \left| 0\right\rangle _1+\beta \left| 1\right\rangle _1)\left|
0\right\rangle _2\rightarrow \left| 0\right\rangle _1(\alpha \left|
0\right\rangle _2+\beta \left| 1\right\rangle _2),
\end{equation}
which describes the QIT from qubit 1 to qubit 2, is completed.

From the description above, it can be seen that the proposal presented here
does not require adiabatic passage (slow variation of the pulse Rabi
frequency), or a second-order large detuning $\delta =\Delta _c-\Delta $
during the entire operation. Here, $\Delta _c=\omega _{32}-\omega _c$ is the
first-order large detuning between the cavity frequency $\omega _c$ and the $%
\left| 2\right\rangle \leftrightarrow \left| 3\right\rangle $ transition
frequency $\omega _{32}$ of the qubits, while $\Delta $ $=\omega
_{32}-\omega $ is the first-order large detuning between the pulse frequency
$\omega $ and the $\left| 2\right\rangle \leftrightarrow \left|
3\right\rangle $ transition frequency $\omega _{32}$ of the qubits. In
addition, one can see that the present proposal does not require a
first-order large detuning $\Delta _c$ or $\Delta ,$ either. Note that only
\textit{resonant} qubit-cavity interaction and resonant qubit-pulse
interaction are used in this proposal, while a second-order large detuning
or adiabatic passage was employed for the previous proposals [6,11-14].
Thus, when compared with the previous proposals [6,11-14], the speed of QIT
in the present proposal is increased by two orders of magnitude.

It can also be seen from the operation above that the present proposal does
not require adjustment of the level spacings of the qubits during the entire
operation, which however was needed by the previous proposal [3].
Furthermore, since the qubit-cavity coupling constants $g_1$ and $g_2$ are
not required to be identical, either nonuniformity in the qubit device
parameters (resulting in nonidentical qubit level spacings) or non-exact
placement of qubits in the cavity is allowed by this proposal.

Several points need to be addressed as follows:

(i) Four levels of each qubit are necessary in order to have qubit 2 (qubit
1) to be decoupled from the cavity mode during the operation of step (i)
[step (ii)].

(ii) The decay of the level $\left| 1\right\rangle $ of each qubit can be
suppressed by increasing the potential barrier between the two lowest levels
$\left| 0\right\rangle $ and $\left| 1\right\rangle $ [1,15,16].

(iii) For simplicity, we considered the identical Rabi frequency $\Omega $
for each pulse during the operations above. However, this requirement is
unnecessary. The Rabi frequency for each pulse can be different and thus the
pulse durations for each step of operations above can be adjusted
accordingly.

(iv) During the operation above, to have the effect of the qubit-cavity
resonant interaction during the pulse negligible, the pulse Rabi frequency $%
\Omega $ needs to be set such that $\Omega \gg g_1,g_2,$ which can be met by
increasing the pulse Rabi frequence $\Omega $ (i.e., by increasing the
intensity of the pulse).

\begin{figure}[tbp]
\includegraphics[bb=51 234 556 677, width=8.2 cm, clip]{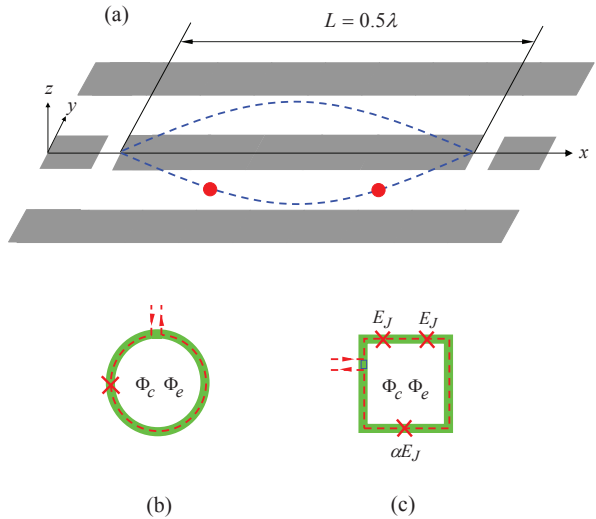} %
\vspace*{-0.08in}
\caption{(Color online) (a) Setup for two superconducting flux qubits (red
dots) and a (grey) standing-wave one-dimensional coplanar waveguide
resonator. $\lambda $ is the wavelength of the resonator mode, and $L$ is
the length of the resonator. The two (blue) curved lines represent the
standing wave magnetic field in the $z$-direction. Each qubit (a red dot)
could be a radio-frequency superconducting quantum interference device (rf
SQUID) consisting of one Josephson junction enclosed by a superconducting
loop as depicted in (b), or a superconducting device with three Josephson
junctions enclosed by a superconducting loop as shown in (c). $E_J$ is the
Josephson junction energy ($0.6<\alpha <0.8$). The qubits are placed at
locations where the magnetic fields are the same to achieve an identical
coupling strength for each qubit. The superconducting loop of each qubit,
which is a large circle for (b) while a large square for (c), is located in
the plane of the resonator between the two lateral ground planes (i.e., the $%
x$-$y$ plane). For each qubit, the external magnetic flux $\Phi _c$ through
the superconducting loop for each qubit is created by the magnetic field
threading the superconducting loop. A classical magnetic pulse is applied to
each qubit through an \textit{ac} flux $\Phi _e$ threading the qubit
superconducting loop, which is created by an \textit{ac} current loop (i.e.,
the red dashed-line loop) placed on the qubit loop. The pulse frequency and
intensity can be adjusted by changing the frequency and intensity of the
\textit{ac} loop current.}
\label{fig:3}
\end{figure}

\begin{center}
\textbf{IV. POSSIBLE EXPERIMENTAL IMPLEMENTATION}
\end{center}

As shown above, it can be found that the total operation times $\tau $ is
given by
\begin{equation}
\tau =\pi /(2g_1)+\pi /(2g_2)+2\pi /\Omega ,
\end{equation}
which should be much shorter than the energy relaxation time $\gamma
_{3r}^{-1}$ and dephasing time $\gamma _{3p}^{-1}$ of the level $\left|
3\right\rangle $ (note that the level $\left| 1\right\rangle $ or $\left|
2\right\rangle $ has a longer decoherence time than the level $\left|
3\right\rangle $), such that decoherence, caused due to spontaneous decay
and dephasing process of the qubits, is negligible during the operation.
And, the $\tau $ needs to be much shorter than the lifetime of the cavity
photon, which is given by $\kappa ^{-1}=Q/2\pi \nu _c,$ such that the decay
of the cavity photon can be neglected during the operation. Here, $Q$ is the
(loaded) quality factor of the cavity and $\nu _c$ is the cavity field
frequency. To obtain these requirements, one can design the qubits (for
solid-state qubits) to have sufficiently long energy relaxation time and
dephasing time, such that $\tau \ll $ $\gamma _{3r}^{-1},\gamma _{3p}^{-1};$
and choose a high-$Q$ cavity such that $\tau \ll \kappa ^{-1}.$

For the sake of definitiveness, let us consider the experimental possibility
using two identical superconducting flux qubits coupled to a one-dimensional
coplanar waveguide transmission line resonator [Fig.~3(a)]. Without loss of
generality, let us consider $g_1\sim g_2\sim 3.0\times 10^9$ s$^{-1},$ which
is available at present [20]. By choosing $\Omega \sim 10g_1$, we have $\tau
\sim 1$ ns, much shorter than $\min \{\gamma _{3r}^{-1},\gamma
_{3p}^{-1}\}\sim 1$ $\mu $s [1,2]. In addition, consider a resonator with
frequency $\nu _c\sim 3$ GHz (e.g., Ref.~[21]) and $Q\sim 2\times 10^4$, we
have $\kappa ^{-1}\sim 1.1$ $\mu $s, which is much longer than the operation
time $\tau $ here. Note that superconducting coplanar waveguide resonators
with a (loaded) quality factor $Q\sim 10^6$ have been experimentally
demonstrated [22,23].

It should be mentioned that the two qubits can be addressed by pulses
separately, through the \textit{ac} current loops placed on their own
superconducting loops [Fig.~3(b,c)]. Note that for superconducting qubits
located in a microwave resonator, the qubits can be well separated, because
the dimension of a superconducting qubit is 10 to 100 micrometers while the
wavelength of the cavity mode for a microwave superconducting resonator is 1
to a few centimeters [6,20]. As long as the two qubits are well separated in
space [Fig.~3(a)], the effect of ``loop current of one qubit'' on the other
qubit and the direct coupling between the two qubits are negligible, which
can be reached by designing the qubits and the resonator appropriately
[6,20]. We should mention that further investigation is needed for each
particular experimental setup. However, this requires a rather lengthy and
complex analysis, which is beyond the scope of this theoretical work.

\begin{center}
\textbf{V. CONCLUSION}
\end{center}

We have presented a way to implement quantum information transfer with
four-level superconducting flux qubits in cavity QED. As shown above, this
proposal avoids most problems existing in the previous proposals, and the
speed of the information transfer is significantly increased when compared
with the previous proposals. The method presented here is quite general,
which can be applied to other physical qubit systems such as atoms and
quantum dots within cavity QED.

\begin{center}
\textbf{ACKNOWLEDGMENTS}
\end{center}

This work was supported in part by the National Natural Science Foundation
of China under Grant No. 11074062, the Zhejiang Natural Science Foundation
under Grant No. Y6100098, the Open Fund from the SKLPS of ECNU, and the
funds from Hangzhou Normal University.

\end{document}